# Integrating urban digital twins with cloud-based geospatial dashboards for coastal resilience planning: A case study in Florida

Changjie Chen[1, *], Yu Han[2], Andrea Galinski[3], Christian Calle[1], Jeffery Carney[1], Xinyue Ye[4], Cees van Westen[2]

1. Florida Institute for Built Environment Resilience, University of Florida, FL, USA
2. Faculty of Geo-Information Science and Earth Observation (ITC), University of Twente, Netherlands
3. Department of Landscape Architecture, University of Florida, FL, USA
4. Department of Landscape Architecture and Urban Planning, Texas A&M University, TX, USA

**Abstract**
Coastal communities are confronted with a growing incidence of climate-induced flooding, necessitating adaptation measures for resilience. In this paper, we introduce a framework that integrates an urban digital twin with a geospatial dashboard to allow visualization of the vulnerabilities within critical infrastructure across a range of spatial and temporal scales. The synergy between these two technologies fosters heightened community awareness about increased flood risks to establish a unified understanding, the foundation for collective decision-making in adaptation plans. The paper also elucidates ethical considerations while developing the platform, including ensuring accessibility, promoting transparency and equity, and safeguarding individual privacy.

## 1. Introduction

Driven by the increasing frequency and intensity of extreme weather events, enhancing resilience in coastal communities has emerged as a crucial imperative. Many coastal communities are also simultaneously grappling with increasing growth and development pressures, which thereby exacerbates the vulnerability of people and infrastructure in harm's way. The disruption caused by these natural hazards can also contribute to the escalation of socio-economic disparities. The U.S Gulf Coast—which face some of the most rapid rates of sea level rise in the world—is experiencing these pressures in particularly acute way with growing hurricane impacts, more frequent "sunny-day" tidal flooding, as well as shrinking flood insurance markets. In fact, the recent Hurricane Idalia was the eighth major hurricane to impact the Gulf Coast in the last six years, and which may bring many communities to a "tipping point" and catalyze new perspectives about future development. In light of these pressing circumstances, it is critical to foster a shared understanding of the risks posed by climate-related hazards and to engage all stakeholders in collective decision-making.

U.S. agencies such as the Federal Emergency Management Agency (FEMA) and the National Oceanic and Atmospheric Administration (NOAA) have spearheaded the development of a variety of tools and frameworks to address the growing demand for effectively managing evolving climate risks. For example, HAZUS, developed by FEMA, is a desktop software for estimating physical, economic, and social impacts of hurricanes and other natural disasters. The National Risk Index, another product of FEMA, provides a framework for categorizing community risks (in five classes) based on social vulnerability and resilience score. The Sea Level Rise Viewer, developed by NOAA, is a web mapping tool to visualize impacts from coastal flooding. However, the majority of these tools focus on analysis and data dissemination at a broader scale instead of facilitating communications among stakeholders at the local community level.

The outbreak of COVID-19 stimulated developments of real-time dashboards to visualize affected communities and provide researchers and policymakers with user-friendly tools to track the pandemic (Dong et al., 2020). The applications of these dashboards have demonstrated a significant potential for integrating datasets from diverse sources into a easily accessible digital platform (Kuster et al., 2023). An urban digital twin (UDT) is defined as a digital counterpart of the physical environment that supports decision-making through the seamless integration of a myriad of geospatial data and analytics techniques. It offers a data-driven platform to visualize, simulate, and analyze interconnected urban systems through a digital replica of the "real" built environment (Batty, 2018; Ye et al., 2022). Due to the interactive and virtual reality features, UDT represents a transformative opportunity for co-creating resilient planning and design strategies, by empowering both policymakers and community members to explore complex relationships, system dynamics, and possible futures across wide-ranging spatial and temporal scales. Therefore, it has drawn much attention in the fields of communicative planning and generative design (Krish, 2011).

This paper investigates the integration of a UDT with a geospatial dashboard, explores the potential of marrying these technologies for community engagement in the context of coastal resilience planning. A geospatial dashboard, alternatively referred to as an urban dashboard or city dashboard, is a web-based interactive interface that leverages spatial analysis, mapping, and graphical summaries to offer a comprehensive overview and key insights about cities (Jing et al., 2019; Kitchin & McArdle, 2016). It has gained significant popularity in various industries including business, healthcare, education, and other government activities. Some well-known examples in the urban realm are the Dublin Dashboard and the London CityDashboard (McArdle & Kitchin, 2016). Dashboards frames the key data narratives concerning the region, while UDTs offer a geospatial representation with unprecedented granularity and specificity that people can easily connect to (Nelson et al., 2022). The synergy of these two technologies can amplify the efficacy of data storytelling, achieving heightened citizen agency in decision-making by bridging the pivotal "last mile of analytics" (Stackpole, 2020).

## 2. Literature Review

The growing necessity to incorporate geospatial data on the vertical scale has led to increased attention in UDTs due to their potential to create visually attractive urban designs using multi-source information and meanwhile embed climate change into planning and management (Pluto-Kossakowska et al., 2022). The concept of UDTs has evolved from their early applications in aerospace and manufacturing to encompass urban environments, including coastal cities (Batty, 2018; Schrotter & Hürzeler, 2020). The main task of urban digital twin is to integrate a variety of data in different formats into a single model. Given increasing size of geospatial datasets, it could be challenging to create and share these datasets efficiently to support urban planning and management. Innovative deep learning methods and geospatial visualization could be used to extract information from images that are reliable and for speeding up UDT creation and updating. Challenges and problems in digital twin need to be solved including sensor data integration issues, scalability and transferability, implementation in practice by communicating with stakeholders, how to connect digital twin with metaverse, what standards to apply. Web-based data visualization interface is another solution to provide an effective model for data-driven analysis and achieve evidence-based decision-making (Praharaj et al., 2023).

In the context of resilience planning, UDTs emerge as a timely and viable approach to informing infrastructure-related policy decisions by incorporating data from crowdsourcing, remote sensing, and social sensing data (Ye et al., 2023). UDTs of critical infrastructure are instrumental in data-driven decision-making for infrastructure management, as they seamlessly connect various datasets and applications. Schrotter and Hürzeler (2020) and Dembski et al. (2019) exemplify this by showcasing the capability of 3D models in transforming themes of cities, including buildings, infrastructure, and vegetation, into the digital world and support participatory planning. By providing real-time and data-rich representations of cities, UDTs empower urban planners and decision-makers to assess vulnerabilities and formulate adaptive strategies effectively. UDTs enable the simulation of various climate change scenarios, helping cities anticipate and prepare for extreme weather events, sea-level rise, and other climate-related challenges (Riaz et al., 2023; Wu et al., 2023). For example, de Vries et al. (2022) utilized a 3D model to visualize earthquake risk and infrastructure vulnerability indicators. Kumar et al. (2018) present a prototype UDT framework for interactive visualization of different flood simulations and multi-stakeholder engagement. Through these simulations, planners can identify areas at risk, evaluate the impact on infrastructure, and optimize mitigation and adaptation measures.

The increased availability of geospatial data stimulated a variety of science-informed planning process to reduce these vulnerabilities and achieve better socially and environmentally intelligent development. The essence of urban planning is to create aesthetically pleasing and socially inclusive spaces for the well-being of citizens. Emerging urban planning concepts, such as sustainable cities, smart growth, and climate resilience, have increasingly focused on individuals' needs (Neuman, 2005; Vella et al., 2016). As a result, human-centered approaches are needed to shift planning focus from technology toward people with more simple solutions (Barton, 2016). UDTs could facilitate this goal through multi-source data integration, augmenting visualization, and stakeholder engagement (Fan et al., 2021). A UDT usually integrates a vast amount of geospatial, environmental, and social data into a single digital representation of the city. This multi-source data integration allows urban planners and designers to have a holistic view of the urban environment, considering not only physical infrastructure but also social and cultural aspects. Wolf et al. (2022) utilized a cloud system to integrate multiple real-time traffic and weather data into a digital twin to facilitate multiple stakeholders' responses to urban flood incidents on roads. As one of the fundamental techniques in UDT, 3D city models are increasingly being used for augmenting visualization of urban infrastructure and environmental planning. 3D city models could enhance visualization and meanwhile enable planners and citizens to explore different design scenarios and visualize how changes will impact the aesthetics and functionality of the city.

The development of 3D visualization tools and functionalities directly promotes applications of the 3D city modeling (Uggla et al., 2023). Most 3D city models adhere to the concepts of level-of-detail (LOD) structure, optimized for view-dependent, and out-of-core rendering to manage massive 3D information (Shirinyan & Petrova-Antonova, 2022). For example, as one of the most important international standards for modeling 3D built environments, CityGML serves as a data exchange format with four levels of detail for 3D visualization and situational awareness purposes in the UDTs (Alva et al., 2022). On the application level, Cesium is a widely used open platform for 3D geospatial data applications (Dimitrov & Petrova-Antonova, 2021). Its 3D tiling pipeline supports various data formats, including point clouds, photogrammetry, 3D city models, imagery, and terrain. Cesium also has an open-source JavaScript library for creating interactive 3D web applications. Multiple web-based applications have been developed relying on Cesium. Hillmann et al. (2022) used Cesium to assess the visual quality of their developed mesh algorithms for large-scale environments.

Nevertheless, to foster collaborative decision-making and engage multiple stakeholders in coastal resilience planning, effectively managing huge 2D and 3D datasets from multiple sources needs to be carefully considered. To facilitate communication among different stakeholders, web-based UDT dashboards are excellent tools for transferring information between a city and its citizens. Negulescu et al. (2023) argue that web-based platforms offer a solution to the challenge of integrating diverse data sources and enabling real-time interactions among stakeholders. By utilizing dynamic and interactive visual analytics, UDT tools serve as virtual meeting spaces where urban planners, policymakers, and residents can actively participate in discussions related to coastal infrastructure and resilience (Kitchin, 2016). They could also provide current and future urban landscapes, and critical infrastructure of cities, and enable planners and policymakers to communicate and educate multiple stakeholders in the planning process. Dembski et al. (2020) developed a prototype UDT platform of Herrenberg, Germany by integrating an urban built environment, street network based on space syntax, and urban mobility simulation and conducted surveys of participants that are involved in the participatory process for mitigation planning of environmental pollution through emission and noise. The survey results indicate virtual interactive UDT models could have a positive impact on participatory planning among different social groups.

## 3. Proposed Framework

This section is dedicated to the realization and implementation of integrating UDT into dashboards. While a broader examination of ethical issues from locational privacy in cloud-based dashboards would be necessary to advance ethical, empathic, and equitable geospatial applications (Nelson et al., 2022), it's worth noting that significant deliberation has been given in addressing a crucial ethical matter during the conceptualization and technology pathway selection of this study. The adoption of decentralized and containerized technologies enables us to leverage privacy protection with minimum cost. Our focus has been on ensuring the accessibility and openness of the proposed development framework, with an emphasis on its feasibility in small and under-resourced communities. This commitment to inclusivity is integral to the ethos of our framework. Generally speaking, the framework encompasses four distinct technical processes (Figure 1): the creation of a 3D city environment, the visualization of large and heterogenous geospatial data, and the development of interconnected "components" in a dashboard (also known as dashboarding), and web deployment.

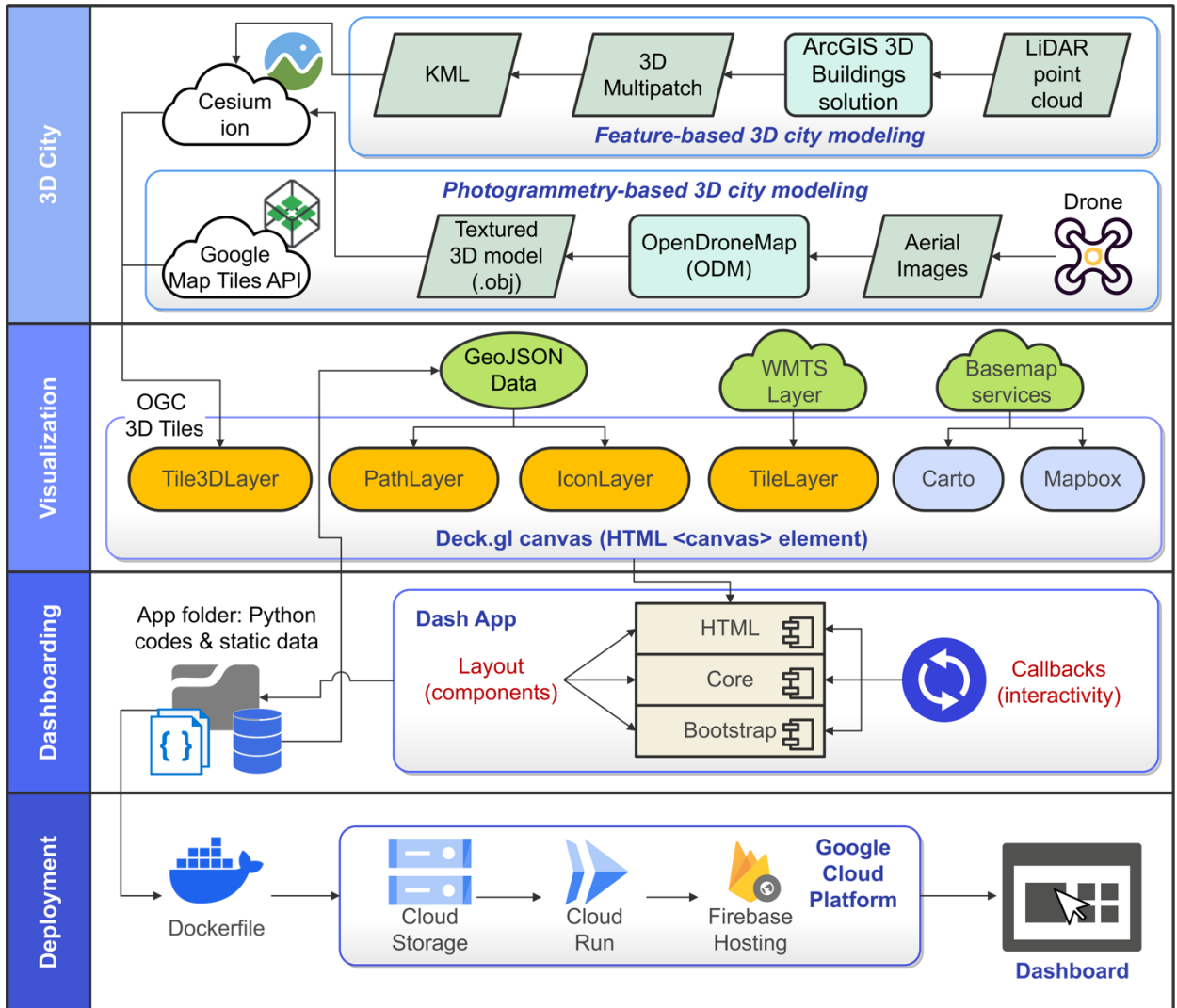

*Figure 1. Overall architecture of the proposed framework.*

### 3.1 Create 3D City Models

Developing a comprehensive UDT relies fundamentally on the creation of accurate and high-fidelity 3D city models (Adreani et al., 2022; Jovanović et al., 2020). In the context of 3D city modeling, a widely adopted measure to indicate model complexity is the Levels of Detail (LOD) classification introduced by the CityGML standard (Chaturvedi et al., 2019; Gröger & Plümer, 2012; Yao et al., 2018). The proposed feature-based 3D city modeling achieves LOD2, in which structural attributes of roofs remains identifiable in individual building's volumetric representation but smaller details (e.g., bulges, dents, window, sills) are neglected (Kolbe et al., 2021). In CityGML 3.0, LOD4 has been dropped, making LOD3 the highest level of detail (Kutzner et al., 2020). However, achieving LOD3 such as the 3D models for the City of Zurich requires tremendous investment in terms of person hours. This presents a significant challenge for small communities to successfully develop and maintain (Schrotter & Hürzeler, 2020). For the purpose of this study and that of other planning related tasks, the abstraction within LOD2 is arguably more appropriate in that it is detailed enough to easily distinguish individual buildings while remains opaque in certain aspects to protect houseowner's privacy. We also introduce an alternative photogrammetry-based approach to visualize a city in 3D, which offers a photorealistic environment to deliver highly immersive user experience. While identifying individual building's information is impossible through this approach, it can be circumvented by overlaying a partially or even fully transparent GeoJSON layer that contains detailed building information.

### 3.1.1 Feature-based 3D City Modeling with LiDAR Data

Recent increase in the availability of Light Detection and Ranging (LiDAR) data in the U.S. and worldwide attracts many research and software companies to improve the process of creating 3D models from such data (Niță, 2021; Shahat et al., 2021; Xue et al., 2020). In 2022, the 3D Elevation Program (3DEP) of the U.S. Geological Survey (USGS) covers 89% of Continental United States (CONUS) and is expected to complete nationwide LiDAR acquisition by 2026 (USGS, 2023). Additionally, the high degree of automation in this process renders it to be a favorable solution to create models at scale over a short period of time. We used ArcGIS Pro and the ArcGIS 3D Buildings Solution to convert raw LiDAR data to ESRI 3D multipatch geometries (ESRI, 2008). The process comprises several steps, including point cloud classification, the generation of raster elevation data for the ground surface, and the creation of 2D layers that serve as the foundation for 3D geometry while incorporating object attributes such as county and municipality information, geometry, and hazard vulnerability data. The initial step, point cloud classification, utilizes ArcGIS 3D Analyst to categorize the core elements of the 3D model, namely the ground and building rooftops. Subsequently, the ground point elevations are employed to construct a digital elevation model (DEM) raster surface, which defines the baseline elevation for each 3D building. Following this, the rooftop points are utilized to generate the building footprints. This procedure involves the transformation of points into a raster surface, followed by their conversion into vector polygons. There is also a manual refinement process to improve the geometries and accuracy of building footprint edges. Finally, the creation of multipatch 3D buildings involves the construction of a Triangulated Irregular Network (TIN) that uses the rooftop points from the point cloud, overlaying the building footprints (ESRI, 2023).

Due to the large volume of 3D data involved in a UDT, especially considering the output 3D models have to be visualized through a browser, it is not recommendable to load and store all 3D geometries on the client side (Botín-Sanabria et al., 2022). Cesium ion (https://ion.cesium.com) can convert large-scale 3D models into 3D Tiles, an OGC standard, that allows users to stream 3D geospatial data hosted in the cloud and only render those within the current extent of the viewport (Cesium, 2023). Cesium ion accepts 3D tiling for data in the CityGML and KML/COLLADA formats and can store per-building properties, provided that the data for each building is a separate model. To convert 3D multipatch geometry to a KML file of separate geometries for individual buildings, we used the Feature Manipulation Engine (FME) Workbench, a desktop software for creating automated workflows that transform data between different formats.

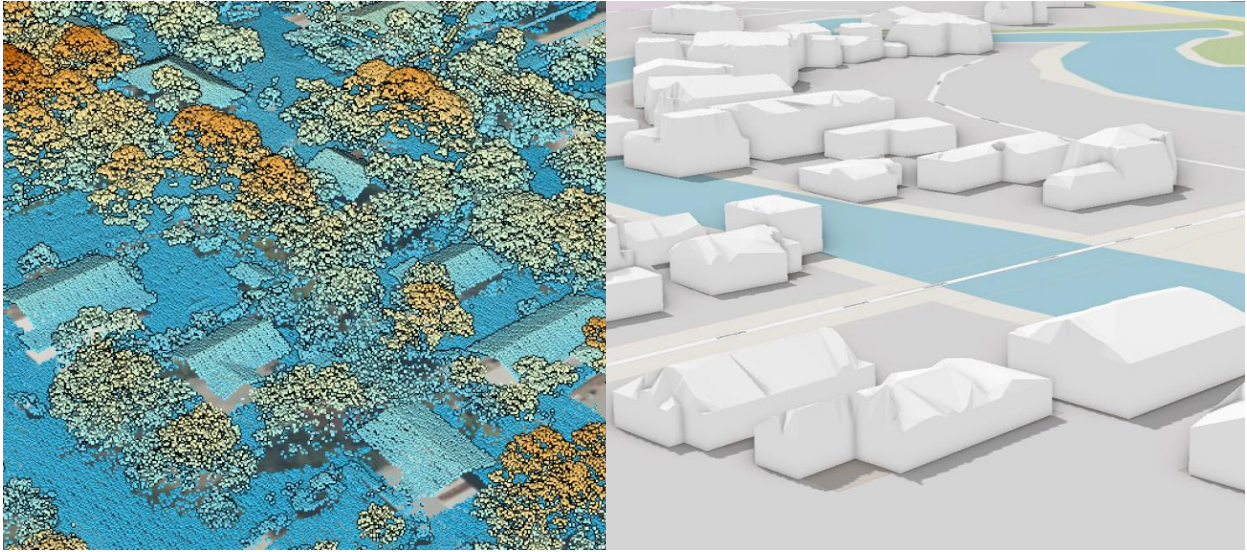

*Figure 2. LiDAR data converted to 3D multipatch geometries. Left: raw LiDAR data (source: https://usgs.entwine.io/). Right: resulting models after the conversion visualized in ArcGIS.*

### 3.1.2 Photogrammetry-based 3D City Modeling

We introduce two pathways to create a photogrammetry-based 3D city environment. The first approach involves making requests via an Application Programming Interface (API), specifically, the Map Tiles API of the Google Maps Platform. Through a token-based mechanism, users can directly access to Google Map's 2D tiles, Street View, and, importantly to UDT development, the Photorealistic 3D tiles. Following the OGC 3D Tiles standard developed by Cesium, the Photorealistic 3D tiles (Figure 3), are essentially 3D meshes textured with high resolution imagery (Google, 2023). It is worthwhile to note that this newly released feature is currently only available to populous urban areas.

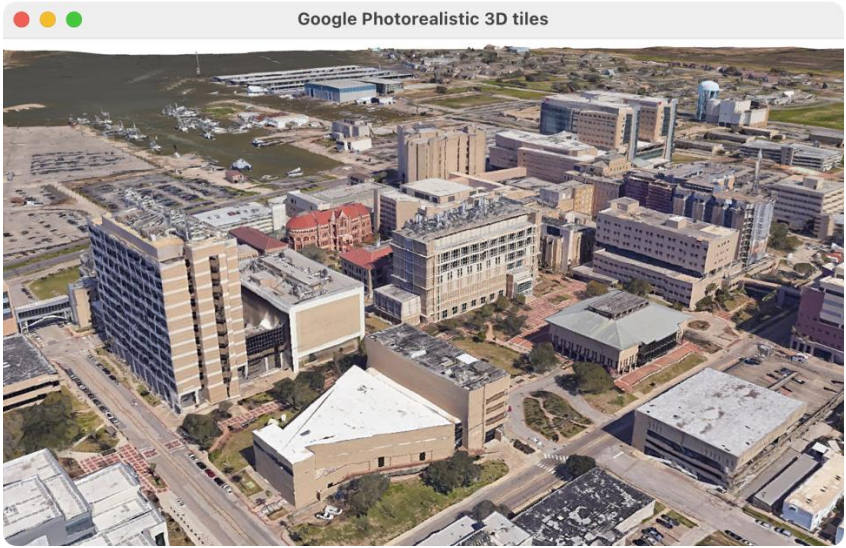

*Figure 3. Photorealistic 3D tiles for UTMB campus in Galveston, Texas rendered in a web browser.*

The second approach involves capturing images from above the area of interest (AOI) using an Unmanned Aerial Vehicle (UAV), i.e., a drone, and postprocessing those images using specialized software. To ensure complete and full coverage of the AOI and to achieve a higher accuracy in the resultant 3D models, a mission planning and automation software is usually used to route the drone's flight path and control its gimbal. Plenty of choices are available for this including proprietary ones (e.g., DroneDeploy, DJI Terra, Site Scan for ArcGIS) and open-source platforms (e.g., Mission Planner, QGroundControl). The technical details of using drones as a remote sensing platform is outside the scope of this article, but Noor et al. (2018) offered a comprehensive review of this emerging technology and the newfound opportunities it brings to urban studies. To create 3D models using collected aerial images, this study utilized OpenDroneMap (ODM), an open-source ecosystem for processing and analyzing drone images (ODM Authors, 2023). ODM can generate 3D textured models in the OBJ format, one of the most popular interchange formats for 3D graphics (McHenry & Bajcsy, 2008). While the output OBJ files do not inherently support a coordinate reference system (CRS), the integration with Cesium ion introduces a georeferencing step during the tiling process. It allows Cesium ion to host these OBJ files in the cloud as 3D Tiles similar to the KML files for the feature-based 3D models. This approach serves as a crucial complement to directly using Google's Map Tiles API, notably in its ability to facilitate the creation of photorealistic 3D environment for rural or less populated urban areas. Additionally, since the on-demand nature of this process, the resulting 3D city are up to date, leading to a more accurate representation of the physical environment. Figure 4 shows an example 3D model in WebODM, a web-based user interface for ODM running locally or in a server.

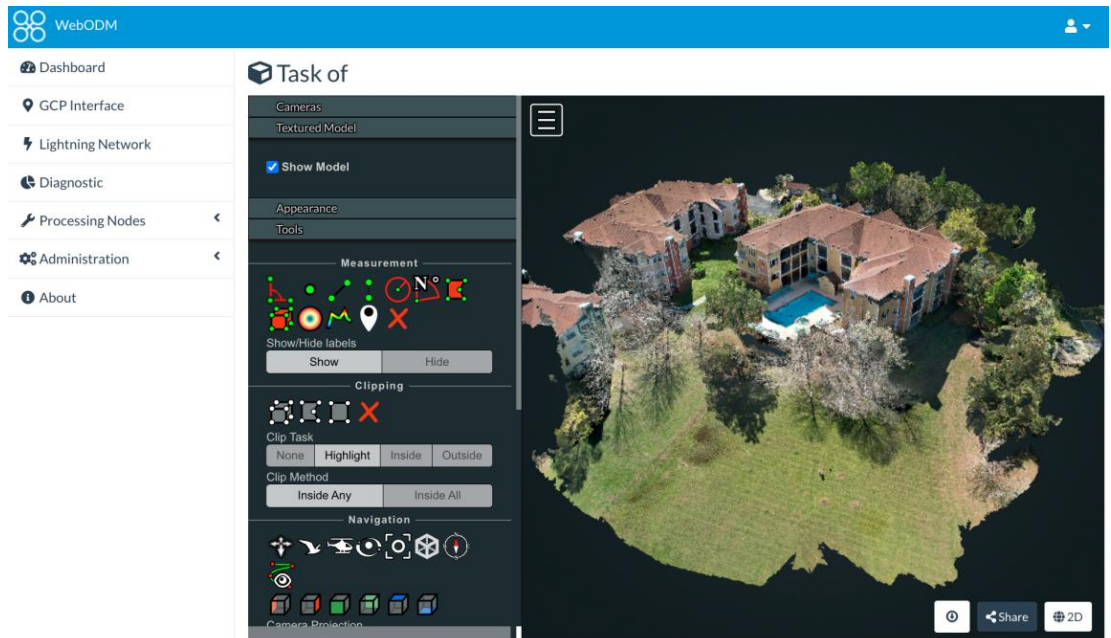

*Figure 4. 3D model of an apartment building and its surrounding features generated by ODM.*

### 3.2 Render Geospatial Datasets Using Deck.gl

Visualizing a web-based UDT must adequately address the heterogeneity and voluminous nature of the datasets while upholding high-performance standards. Specifically, different features and phenomena within the physical environment are inherently varied in their representation in the digital environment, which underscores the importance of addressing data interoperability as a fundamental challenge and a key consideration in the development of UDTs (Petrova-Antonova & Ilieva, 2019). Besides 3D city models, a UDT could include a digital terrain model (DTM), raster tiles served through the Web Map Tile Services (WMTS) protocol, incoming JavaScript Object Notation (JSON) data synchronized with sensors in real time, and simple vector geometries (e.g., points, lines, and polygons) in the GeoJSON format (Rantanen et al., 2023). Hence, when selecting the underlying visualization framework for online, public-facing UDTs, developers must prioritize the seamless coexistence of these heterogeneous datasets within the web environment, ensuring that the display of one dataset do not impede or compromise the presentation of others. Additionally, the magnitude of such datasets in a UDT can be substantial, reaching potentially millions of data points (Lei et al., 2023). Considering the need for interactivity, it's likely that these datasets will also undergo frequent changes, both in terms of their position and style.

After scrutinizing multiple existing visualization frameworks (Table 1), we found Deck.gl (https://deck.gl/) is most versatile and suitable, especially for developing web-based UDTs. As an open-source framework, Deck.gl is designed to focus on scalability, usability, and extensibility (Wang, 2019). In a standard visualization pipeline (Figure 5), Deck.gl employs General-Purpose Graphics Processing Units (GPGPU) for all computations, as opposed to a Central Processing Unit (CPU). This approach enables a significant degree of parallelism, thereby achieving minimal latency in the visualization process. The framework was originally developed by a team at UBER Technologies, Inc. to evaluate and oversee the functionality of their system. Consequently, it has been designed with a focal point on geospatial visualization, exemplified by supporting the Mercator projection, transformation of data from a Latitude and Longitude-based representation in degrees to a consistent Cartesian space, where uniform units are applied across all three dimensions. Deck.gl is highly portable and customizable. It seamlessly integrates with various web frameworks, including React, while also accommodating usage in Vanilla JavaScript. Moreover, it also offers a Python binding, specifically Pydeck, ensuring adaptability to a wide array of web architectures.

*Table 1. A comparison on key features of reputable frameworks for web visualization.*

| Name | Open Software | Computation | 3D Support | Geospatial Support | Dynamic Styling | Extensibility |
|---|---|---|---|---|---|---|
| Deck.gl | Yes | GPU | Yes | Yes | Yes | High |
| Three.js | Yes | GPU | Yes | No | No | High |
| CesiumJS | Yes | GPU | Yes | Yes | Yes | Moderate |
| Leaflet | Yes | CPU | Limited | Yes | Limited | High |
| D3.js | Yes | CPU | No | No | Limited | High |
| Mapbox GL JS | Yes | CPU/GPU | Yes | Yes | No | High |
| ArcGIS API for JS | No | CPU/GPU | Yes | Yes | Yes | Moderate |

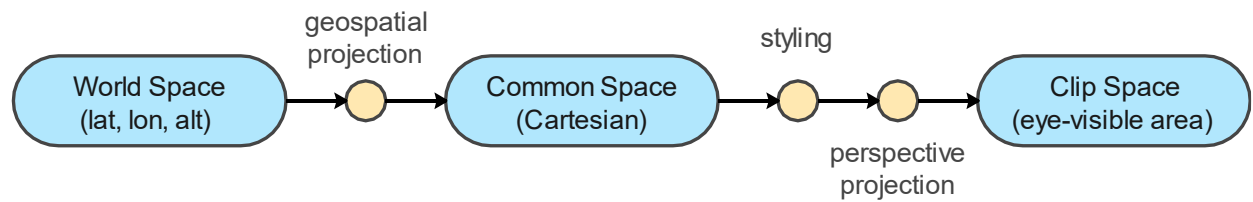

*Figure 5. A process of web-based geospatial visualization.*

In the context of UDT visualization, the Primitive-Instancing-Layering (PIL) paradigm of Deck.gl streamlines the integration and rendering of a diverse array of inputs, encompassing static data, modeling results, and an assortment of storm event scenarios, each with varying time horizons. In this study, we constructed a composite representation of the physical environment by layering four distinct components (Figure 6). Deck.gl's Tile3DLayer takes center stage, seamlessly integrated with 3D data sourced from Cesium ion. This layer further empowers users with the ability to selectively interact with individual 3D models, facilitating interactive exploration and the retrieval of detailed information on a feature-specific basis. In tandem, we harnessed Deck.gl's TileLayer to proficiently render WMTS raster tiles hosted on ArcGIS Online, catering to the visualization requirements of our vulnerability assessment outputs. Lastly, the GeoJsonLayer supplements this versatile ecosystem by enabling the rendering of diverse geospatial entities, ranging from specific assets represented as points, to intricate road networks manifested as paths, and the delineation of parcels or census blocks in the form of polygons. At the bottom of this "deck" of layers, the basemap provides invaluable contextual information about the physical environment, sourced from providers such as Mapbox and Carto, each offering an array of customizable styles, including options like Satellite Imagery and standard Roadmap.

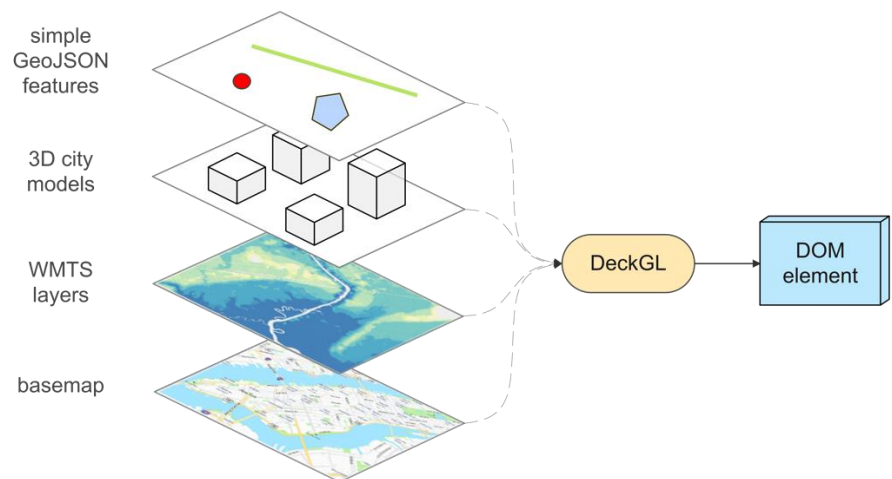

*Figure 6. Deck.gl's layered approach to visualize data from various sources.*

### 3.3 Integrate UDT with Dashboard

In this study, dashboard development was accomplished using Dash (https://plotly.com/dash/), a Python framework for declaratively building full-stack web applications (Hossain, 2019). This approach is notably advantageous for individuals lacking front-end development expertise, such as urban planners, enabling them to craft interactive and engaging dashboards for the effective communication of ideas and demonstration of research findings (Brown et al., 2023). In practice, developers can just declare high-level descriptions concerning the dashboard's structure, styling, and specific responses to user interactions, sparing them the intricacies of low-level implementations. A Dash-based dashboard application contains two fundamental parts: "layout" and "callbacks." The former pertains to specific elements, known as components, within the application, while the latter deals with interactivity, both among multiple components and within a single component.

First, from a "layout" perspective, components of a geospatial dashboard fall into three categories: HTML, Core, and Bootstrap components. Our UDT built with Deck.gl is integrated, as a self-contained HTML component (Figure 6), into the dashboard's Document Object Model (DOM). Within the geospatial dashboard, UDT takes center stage, complemented by multiple infographics. The seamless integration of the UDT with additional visual aids, such as interactive charts, is a pivotal reason for constructing the dashboard with the Dash framework. Created on top of Plotly, Dash natively supports transforming a Plotly figure object into a Graph, a Dash Core component. These graphs offer a comprehensive overview of the extent to which critical assets are affected by flooding under specific scenarios. Moreover, we implemented a multi-page architecture to isolate distinct functionalities of the dashboard application such as home, viewer, and tracker. An organized and visually coherent interface is conducive to the intuitive exploration and effective interpretation of the information within the dashboard. Therefore, we designed the user interface (UI) as a grid system using an extension of Dash, namely Dash Bootstrap components.

Secondly, Dash "callbacks" orchestrates user interactions against individual dashboard components, each of which is defined by a set of keyword arguments, i.e., properties. A callback is a Python function modified by the "@callback" decorator. Through Dash Input and State objects, a callback can update, or recreate, any component (as Dash Output) within the dashboard by changing the component's properties. This approach is commonly referred to as "Reactive programming" since Output objects automatically "react" to changes in the Input objects. While Dash is a Python framework, it offers the flexibility to extend its applications with custom JavaScript. Leveraging this capability, we developed a guided tour using Shepherd.JS (https://shepherdjs.dev/) to enhance the user experience (UX).

### 3.4 Deploy Dashboard Using Google Cloud Platform

Aiming at strengthening community resilience, the value and effectiveness of our platform with integrated UDT and geospatial dashboard hinges on its accessibility to stakeholders and its tangible contribution to the decision-making processes concerning adaptation strategies. This serves as a fundamental motivation underpinning the development of a web-based UDT platform. For the purpose of Continuous Integration and Continuous Deployment (CI/CD), we have harnessed the capabilities of the Google Cloud Platform (GCP) to establish a robust and scalable deployment process. The objective is to minimize the required effort during the deployment process and reducing the associated hosting cost. Initially, we employ Docker to create containerized images, which are subsequently stored in Google Cloud Storage. This containerization not only streamlines deployment but also ensures consistent execution across diverse environments. Following this, we utilize Google Cloud Run for building and deploying our application, harnessing the advantages of serverless computing. This approach guarantees scalability, resource efficiency, and improved reliability. To enhance accessibility and strengthen brand identity of our geospatial dashboard, we further employed Firebase Hosting. This enables us to replace the randomly generated domain name with customized ones, resulting in a user-friendly and memorable uniform resource locator (URL) address ending with "web.app." This deployment strategy on the GCP aligns with best practices in cloud computing for web development, ensuring that our dashboard is scalable, accessible, and user-centric.

### 4. Results

Our proposed framework outlines a technology roadmap to develop UDTs and integrate them into cloud-based geospatial dashboards, aiming to facilitate coastal communities in their adaptation planning for sea-level rise. Within this context, a UDT provides an exceptional visual platform to present outputs of compound flooding models alongside LOD2 3D city models (Figure 7). The results confirm that Deck.gl has the capability to integrate harmonious coexistence of heterogenous sources of input data within the same application. Additionally, the successful implementation underlines its capacity to efficiently handle large dataset visualization in the web environment. Enhanced by per-feature selection capabilities, the immersive environment possessed in UDT naturally engage users to interactively explore and extract relevant information concerning the potential flood risk. The interactivity of our UDT allows for viewing flood maps from various spatial scales, facilitating a nuanced understanding of the potential risks.

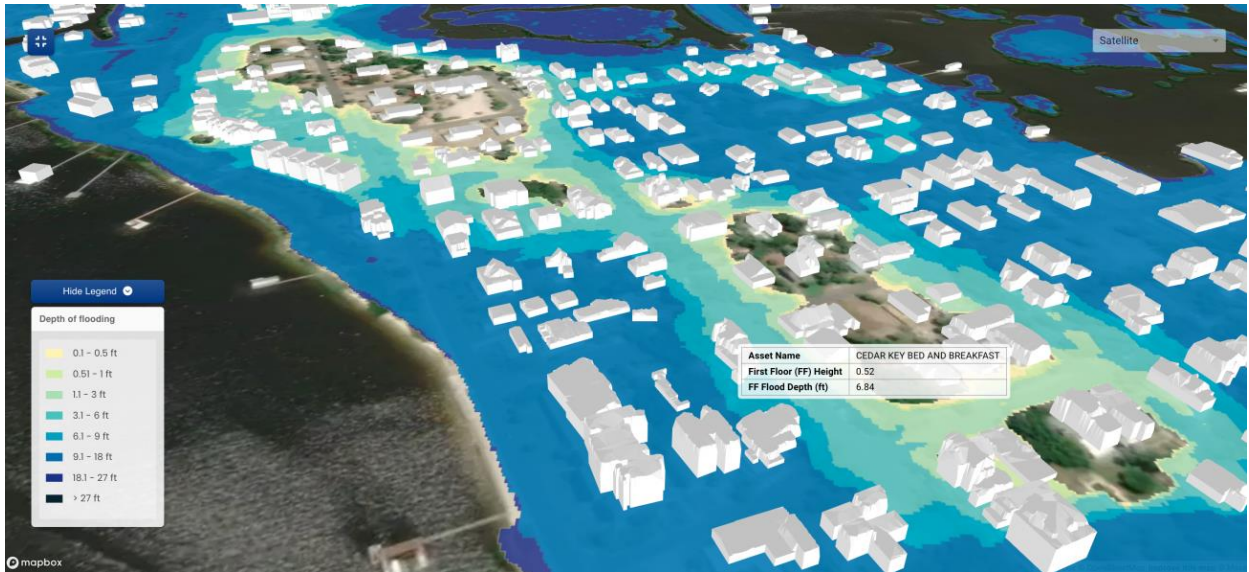

*Figure 7. Visualizing potential flood risk within a UDT.*

Scenario exploration is made possible through a slider mechanism. Utilizing the Dash framework, we included two sliders (Figure 8) in the geospatial dashboard. The horizontal slider consists of three values of time permitting capturing sea-level rise effects, whilst the vertical slider encompasses eight distinct weather conditions. The combination of these two sliders generates a total of 24 scenarios, rendering a holistic picture of the true vulnerability the city is facing presently and in the foreseeable future. Furthermore, the vulnerability assessment categorizes critical assets into different thematic groups. Using the capabilities of the dashboard, users can visualize critical assets particularly of their interests by clicking a button in the navigation bar at the top of the dashboard as shown in Figure 9. In contrast to the conventional practice of displaying printed GIS maps, this "self-learning" approach to public participation demonstrated its effectiveness in raising awareness among residents. This observation was made during multiple public outreach meetings conducted in different localities.

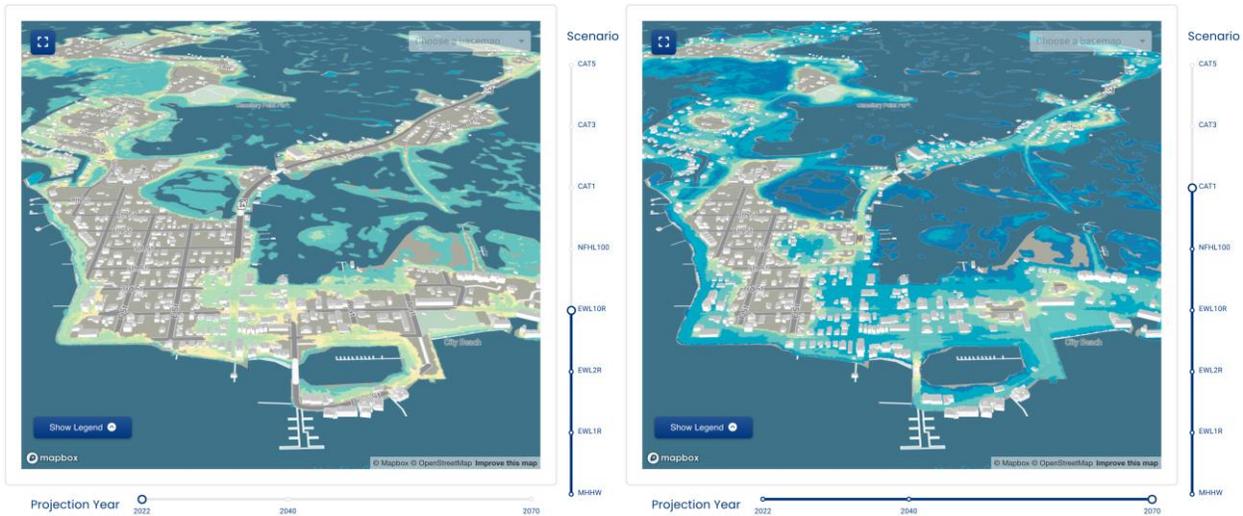

*Figure 8. Using sliders to explore various scenarios. Left: Projected flood map for the extreme water level over a 10-year return period (EWL10R) in 2022 (the start time serving as a baseline of the analysis). Right: Projected flood map for a Category-1 Hurricane in 2070.*

To a substantial extent, the development of UDTs revitalizes opportunities for community engagement in an unprecedented fashion. Powered by 3D visualization and web technology, our UDT functions as a practical educational tool, equipping residents with an in-depth understanding of vulnerability and encouraging informed actions. This heightened awareness plays an important role in fostering a shared perspective on the challenges at hand. Moreover, the platform extends its utility beyond the individual level to offer a means for local municipalities to assess community-wide vulnerability. As shown in Figure 9, decision-makers not only have the ability to pinpoint critical nodes within the city's road network but also are provided with detailed information concerning the percentage of road segments affected by flooding and the degree of that impact in terms of flood depths. From this perspective, the integration of UDT and geospatial dashboard can indeed facilitate in the decision-making process for devising adaptation strategies.

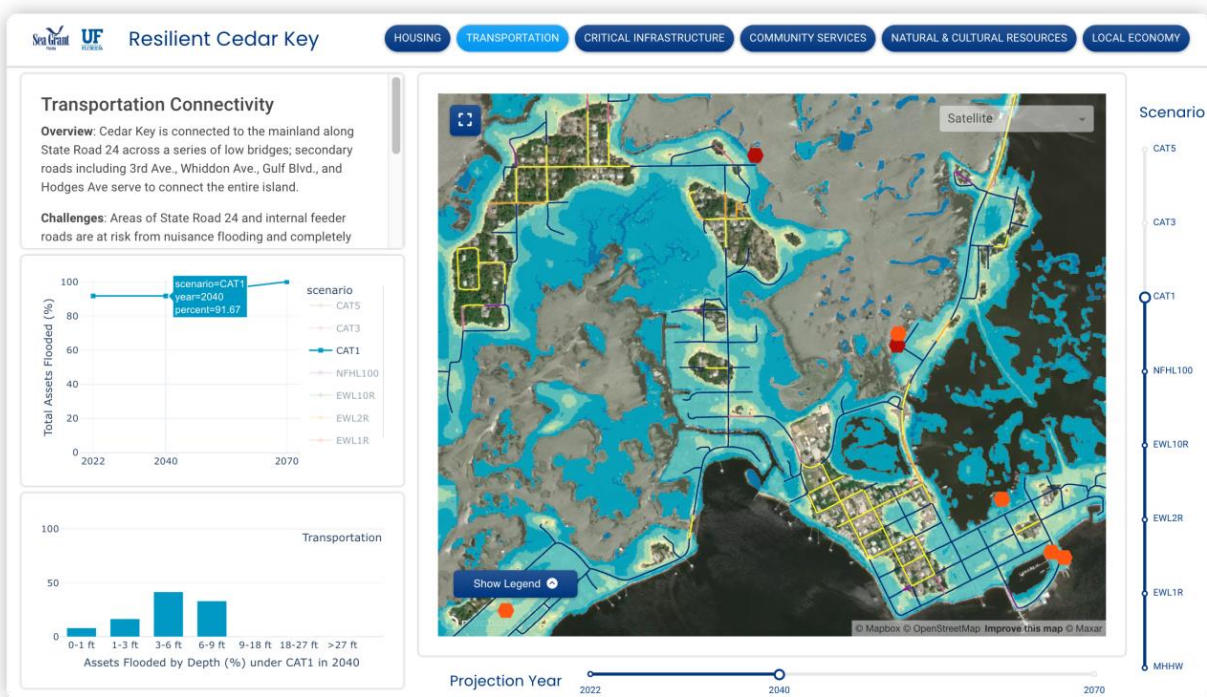

*Figure 9. A static view of the full dashboard.*

## 5. Case Study and Discussion

The Resilient Cedar Key project, funded by the Florida Department of Environment Protection (FDEP), has employed the integrated UDT and dashboard (www.resilientcedarkey.web.app) throughout the community engagement process. The data visualization and interaction were curated for two user groups including a smaller stakeholder workgroup (the Task Force), as well as residents of the town. The Task Force was comprised of representatives for various interests such as the Chamber of Commerce, Water and Sewer District, Public Works, Fire Department, Florida Shellfish Aquaculture Association, Florida Nature Coast Conservancy, UF/IFAS Shellfish Aquaculture Extension, and others. The web application was used for three stakeholder meetings as well as three general public outreach meetings. The first two meetings focused on the results of the vulnerability assessment, while the second four meetings introduced the proposed adaptation action areas and conceptual ideas for associated projects. Lastly, Hurricane Idalia made landfall just north of Cedar Key on August 30, 2023; in the run up to this event, city officials had requested that the team use the web application to show how the island would be affected by 7-foot and 11-foot flood depths, which was the initial projected impacts of a Category 4 Hurricane.

While the community was braced for the worst, the storm was not as severe as expected due to its northward drift. However, this event spurred the use of the tool not only for long-term resilience planning, but also real-time estimates of potential impacts due to tropical storms and hurricanes. Especially given the rapid intensification of several recent hurricanes—Ida (2021), Ian (2022), Idalia (2023), and Hurricane Otis (2023)—quickly illustrating a range of flood scenarios is particularly important for coastal communities as they prepare for an event and its immediate aftermath.

The web application was instrumental in several informational, educational, and communication capacities. It was used to communicate the vulnerability assessment and the adaptation actions to the task force and the general public. As detail-oriented subject matter experts, the task force was keen to learn more about the technical nuances of the analysis and found the dashboard helpful as they could investigate many different flood scenarios (toggling between different flood hazards at different time periods) to determine how these may impact their area of interest (roadways, housing, wastewater facilities, fishing docks, etc.). In particular, it was common for these stakeholders to have interest in exploring the many different smaller, more frequent flood events to understand the more common impacts that may be expected to their homes, businesses and/or industries. Also, there was general agreement by task force members that the vulnerability results were "very complex and a lot to take in," and, for example, a PowerPoint presentation of dozens of maps would not be suitable to present to the public. The web app thus played an important role in the community outreach meetings as the project team could focus on sharing concise, high-level information, but also provide a more in-depth resource for individuals who wanted to learn more. At public events individual workstations were set up where researchers could help community members navigate the tool, often finding their home or exploring flood scenarios that were of interest to them.

In the most recent community workshop in October 2023 (Figure 10), the researchers included a new tab titled "adaptations" which overlayed a series of resilience and adaptation projects. Like the standard flood scenarios, community members could evaluate different proposed projects against different flood scenarios. The seamless step from vulnerability assessment to adaptation plan created a sense of transparency and immediacy to the process. It allowed community members to see for themselves how ambitious adaptation projects (that might require substantial investment) were directly responsive to the vulnerability that their community faced.

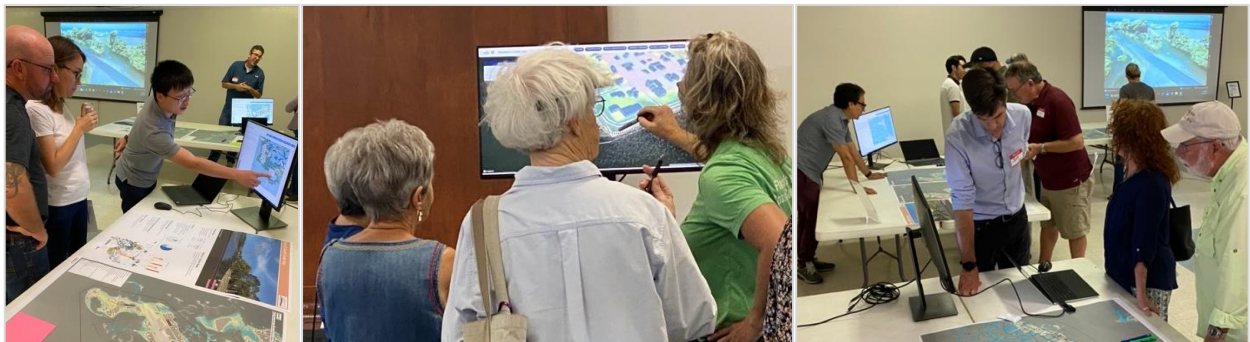

*Figure 10. Usage of the Resilient Cedar Key Dashboard in Public Meetings.*

### 5.1 Ethical Considerations

While developing UDT applications, it is important to adhere to data protection regulations and security standards to safeguard sensitive information of residents. For instance, our approach involves the careful use of GeoJSON layers and partial or fully transparent overlays, which allow the visualization of detailed building data while maintaining the confidentiality of homeowners. Other datasets carry other concerns that can adversely impact individuals including personal health information, home values, and more. It is imperative that ethical consideration at all stages ensure that individual privacy is preserved while providing valuable insights into vulnerability assessment and adaptation planning. Deploying through the Google Cloud Platform, we have not only ensured scalability, accessibility, and user-centricity but have also implemented encryption and secure authentication protocols to facilitate ethical data handling. By following the best practice in cloud computing for web development, such as containerization and serverless computing, we also minimize cost in hosting and maintaining the web application.

The authors contend that for a platform to achieve true accessibility, it must prioritize user-friendliness and intuitiveness, effectively catering to a diverse user base. To achieve this, we have not only implemented a user-friendly interface design to facilitate data storytelling but also enriched the user experience through the integration of a guided tour presented to users upon the initial login. These design implementations play a pivotal role in fostering a self-learning process as users navigate the wealth of the data-rich content within the dashboard. By encouraging a sense of autonomy in information retrieval, residents are empowered to actively engage with the data, thereby nurturing a more profound understanding of the insights presented. This approach acknowledges the efficacy of self-guided learning, bolstering trust and cultivating a shared sense of responsibility within the community. Moreover, such user-centric designs of this platform not only stimulate active participation but also promote transparency in the decision-making processes. These considerations collectively underscore a commitment to ensuring the accessibility and effectiveness of the platform.

## 6. Conclusion

The integrated UDT and dashboard introduced in this paper offers a novel platform for community engagement in the context of coastal resilience planning. For local communities, the developed UDT is not just an integrated technological tool. It also serves as a vital educational tool. Stakeholders from different backgrounds benefit from its ability in visualizing, simulating, and assessing the impacts of climate change in local communities in an accessible and user-friendly manner. The scenario exploration in our platform enables stakeholders to examine various sea-level rise scenarios and assess their vulnerabilities comprehensively and in a comfortable small group or individual setting. The case study demonstrates the usage of this platform in public outreach meetings and workshops, which have underscored its effectiveness in raising awareness and promoting meaningful dialogue between community members, public officials, professionals, and academic planners. The interactive geospatial dashboard contributes to a sense of competency and control that is often lacking for community members. Especially in small communities, rural communities, and post-disaster areas, providing the community with the skills to navigate the system, and inform their own thinking about flood risk has contributed greatly to the success of adaptation planning in the case studies discussed here.

As an emerging technology, UDT's full potential has yet to be fully investigated. The iterative development of the UDT encourages the inclusion of new and developing technologies which provides immediate insight and rapid advancement. During the development of photogrammetry-based 3D city model, this research team discovered that Google Photorealistic 3D tiles are only available for metropolitan areas reaching a certain population density. This excludes small coastal communities the benefit of this cutting-edge technology. Our work aims to address challenges in UDT applications for communities with limited financial and human resources where innovation can provide far greater access to data and visualization. This evolving framework allows for creating dynamic and user-friendly dashboards, enabling interactive exploration and a self-learning approach for educating a broader audience and provides a robust tool to address the multifaceted challenges posed by coastal adaptation in the context of climate change.